\documentclass[10pt,aps,prd,twocolumn,floats,floatfix,showpacs,superscriptaddress,nofootinbib,longbibliography]{revtex4-2}
\usepackage[utf8]{inputenc}               		

\usepackage{calligra}
\usepackage{graphicx,mathtools,amssymb,amsmath,amsthm,amsfonts,epsfig,epsf,multirow}
\usepackage[outdir=./]{epstopdf}
\usepackage[dvipsnames]{xcolor}
\usepackage[linktocpage]{hyperref}
\hypersetup{
    colorlinks=true,
    linkcolor=blue,
    filecolor=magenta,      
    urlcolor=BrickRed,
    citecolor=BrickRed,
}
\hypersetup{pdfstartview=}	
\usepackage{tensor}	
\usepackage{csquotes}
\usepackage{mathrsfs}

\usepackage{comment}

\DeclarePairedDelimiter{\abs}{\lvert}{\rvert}

\newcommand\normL[1]{\left\lVert#1\right\rVert_{L^{2}}}

\newcommand{\mbf}[1]{\mathbf{#1}}

\usepackage{graphicx}
\usepackage{amsmath}
\usepackage{autobreak}
\usepackage{float}

\usepackage{multirow}
\usepackage{color, colortbl}

\usepackage{bm}

\usepackage{lipsum}

\usepackage{soul}

\begin{document}

\author{Farid~Thaalba}
\address{Nottingham Centre of Gravity,
Nottingham NG7 2RD, United Kingdom}
\address{School of Mathematical Sciences, University of Nottingham,
University Park, Nottingham NG7 2RD, United Kingdom}

\author{Nicola~Franchini}
\affiliation{Université Paris Cit\'e, CNRS, Astroparticule et Cosmologie,  F-75013 Paris, France}
\affiliation{CNRS-UCB International Research Laboratory, Centre Pierre Binétruy, IRL2007, CPB-IN2P3, Berkeley, CA 94720, USA}

\author{Miguel~Bezares}
\address{Nottingham Centre of Gravity,
Nottingham NG7 2RD, United Kingdom}
\address{School of Mathematical Sciences, University of Nottingham,
University Park, Nottingham NG7 2RD, United Kingdom}

\author{Thomas~P.~Sotiriou}
\address{Nottingham Centre of Gravity,
Nottingham NG7 2RD, United Kingdom}
\address{School of Mathematical Sciences, University of Nottingham,
University Park, Nottingham NG7 2RD, United Kingdom}
\address{School of Physics and Astronomy, University of Nottingham,
University Park, Nottingham NG7 2RD, United Kingdom}

\begin{abstract}
We study spherical evolution in scalar-Gauss-Bonnet gravity with additional Ricci coupling and use the gauge-invariant approach of Ref.~\cite{Reall:2021voz} to track well-posedness. Our results show that loss of hyperbolicity when it occurs, is due to the behaviour of physical degrees of freedom. They provide further support to the idea that this behaviour can be tamed by additional interactions of scalar. We also point out a limitation of this gauge-invariant approach: the fact that field redefinitions can change the character of the evolution equations.          
\end{abstract}

\title{Hyperbolicity in scalar-Gauss-Bonnet gravity: a gauge invariant study for spherical evolution}
\maketitle
\section{Introduction}
The covet to test general relativity (GR) is at an all-time high~\cite{LIGOScientific:2021sio,Berti:2015itd,Yunes:2024lzm,Colpi:2024xhw,Gair:2012nm,Yagi:2013du,Barack:2018yly}, ignited by the landmark detection of gravitational waves (GWs) by the LIGO and Virgo collaboration \cite{LIGOScientific:2016aoc} and the large number of GW event that followed~\cite{LIGOScientific:2018mvr,LIGOScientific:2020ibl,LIGOScientific:2021usb,KAGRA:2021vkt}. Third-generation detectors, e.g., the Einstein telescope~\cite{Punturo_2010,Sathyaprakash:2012jk}, and space-born detectors such as LISA~\cite{Barausse:2020rsu,LISA:2022kgy,Colpi:2024xhw} will enable the search for possible anomalies in our understanding of the gravitational interaction in an unexplored territory of extreme dynamical gravity.     

The simplest way to test GR against the data from binary coalescences is to use its predictions as a null hypothesis. 
Nonetheless, obtaining waveforms which include deviations from GR has many benefits. It would allow one to interpret the new physics if deviations were to be detected.  It can provide quantitative bounds, with the tightest ones coming from theory-specific modelling. The theory-specific waveforms can also be used to calibrate more generic parametrizations to improve their accuracy.

Suppose we subscribe to the effective field theory (EFT) perspective (see, e.g.,~\cite{Burgess:2020tbq} for a comprehensive introduction). In that case, we can organize our ignorance about the UV theory in a tower of operators, which we can truncate at a specific order. In principle, the coefficients of these operators can be obtained from experiments. In the bottom-up approach of EFTs, one is to identify the relevant low-energy degrees of freedom (dof) and their symmetries (e.g., diffeomorphism invariance, local Lorentz symmetry) and then write all the compatible interactions between the dof that respect said symmetries, and organize these interactions based on a power-counting scheme.

A plethora of extensions of GR or the standard model (SM) of particle physics include a scalar as an additional light dof. These models might address the hierarchy problem (or other naturalness problems in the SM), the accelerated expansion of the universe, or elucidate what dark matter might be~\cite{Copeland:2006wr,Essig:2013lka,Hui:2016ltb,Barack:2018yly}. If such a scalar couples to curvature it can leave an imprint on extreme gravity dynamics. The Horndeski action~\cite{Horndeski:1974wa} can be seen as an EFT in this setup.

One notable term within the Horndeski class is a coupling of the scalar to the  Gauss-Bonnet (GB) invariant $\mathcal{G}=R^{\mu \nu \rho \sigma} R_{\mu \nu \rho \sigma}-4 R^{\mu \nu} R_{\mu \nu}+R^{2}$, where $R_{\mu \nu \rho \sigma}$, $R_{\mu \nu}$, and $R$  being the Riemann tensor, Ricci tensor, and Ricci scalar respectively. Such an interaction is known to endow black holes with an additional scalar charge~\cite{Kanti:1995vq,Yunes:2011we,Sotiriou:2013qea,Sotiriou:2014pfa}. Black holes in this class of theories can also ``spontaneously scalarize" (i.e., the scalar field grows and dresses the horizon with a non-trivial configuration) if the curvature is sufficiently high near the horizon~\cite{Silva:2017uqg,Doneva:2017bvd}, or if they rotate rapidly enough~\cite{Dima:2020yac,Herdeiro:2020wei,Berti:2020kgk} (see~\cite{Doneva:2022ewd} for a review). 

A thorny issue for models including this interaction term, and for alternative theories more broadly, is setting up a well-posed initial value problem, i.e.~the solution to the system of partial differential equations (PDEs) should be unique and depend continuously on the initial conditions \cite{hadamard}).
The character of a PDE system can usually be checked by examining its principal symbol (containing the coefficients of the highest derivative terms). For an initial value (or Cauchy) problem (IVP), the character of the system must be strongly hyperbolic; that is, the principal symbol is diagonalizable with real eigenvalues. We review these notions in some detail in appendix~\ref{sec:hyperbolicity}. 

 In the context of scalars nonminimally coupled to gravity, and for the coupling to the GB invariant in particular, early attempts to study nonlinear evolution have resorted to working perturbatively in the coupling constant to circumvent well-posedness issues~\cite{Benkel:2016rlz,Benkel:2016kcq,Okounkova:2017yby,Witek:2018dmd,Okounkova:2019dfo,Silva:2020omi, Elley:2022ept}. This approach loses accuracy for longer evolution times due to secular growth effects~\cite{GalvezGhersi:2021sxs}, which can also drive it outside its range of validity. Additionally, it cannot capture nonlinear effects within the scalar sector.   Another proposal draws inspiration from the treatment of viscous relativistic hydrodynamics in GR and tames the (spurious) behaviour of certain dofs to ``fix" the equations and ensure well-posedness~\cite{Cayuso:2017iqc,Allwright:2018rut,Cayuso:2020lca,Franchini:2022ukz,Cayuso:2023aht,Lara:2021piy,Bezares:2021yek,Cayuso:2023aht,Gerhardinger:2022bcw,Lara:2024rwa,Corman:2024cdr}. Well-posed formulations without any approximations or ``fixing" have been shown to exist when the theory remains weakly coupled, i.e., the beyond GR effects remain ``small" compared to GR then a well-posed formulation exists \cite{Kovacs:2020ywu}. How far one can push the couplings for particular initial data before such a formulation fails is hard to assess. Several numerical studies of well-posedness concerning scalar-Gauss-Bonnet theories have been conducted in the literature~\cite{Ripley:2019hxt,Ripley:2019irj,Ripley:2020vpk,East:2020hgw,East:2021bqk,Corman:2022xqg,AresteSalo:2022hua,R:2022hlf,Corman:2024vlk,Corman:2024cdr,Lara:2024rwa}. 
 
To keep calculations tractable and limit the size of the parameter space, most of the numerical explorations have included only the scalar-GB interaction. However, in an EFT, one would expect other terms to be present as well, and they could influence hyperbolicity. Indeed, it was shown recently that the addition of a coupling between the scalar field and the Ricci scalar can be crucial for well-posedness~\cite{Thaalba:2023fmq,Thaalba:2024htc} in spherical evolution. These two studies used different gauges.
In~\cite{Reall:2021voz}, a gauge-invariant diagnostic tool for hyperbolicity in scalar-tensor theories was introduced.
Here, we use this gauge-invariant approach to study hyperbolicity and loss of well-posedness for spherical evolution in scalar Gauss-Bonnet gravity with an additional Ricci coupling. 

As we will show, in our setup, the gauge-invariant hyperbolicity reduces to tracking the signature of an ``effective metric". We do that in two different gauges and for various coupling constants and initial data. We find that the determinant of the effective metric changes sign when hyperbolicity is lost consistently with other diagnostic tools (see, e.g.,~\cite{Thaalba:2023fmq,Ripley:2019hxt,Ripley:2019aqj}). Our results elucidate the relation between the results of~\cite{Reall:2021voz} and~\cite{R:2022hlf}. They demonstrate that the loss of hyperbolicity during spherical evolution in scalar Gauss-Bonnet gravity is not just due to a bad gauge choice, but due to the behaviour of physical degrees of freedom that can indeed be tamed by the presence of additional interactions --- in our case the Ricci coupling. Finally, we examine the effect of field redefinitions on hyperbolicity and point out that this constitutes a limitation of using gauge-invariant criteria for well-posedness.

The paper is organized as follows: in section~\ref{sec:PS_char}, we discuss some results concerning the principal symbol and the characteristics of second-order scalar-tensor theories. In section~\ref{sec:theory}, we focus on scalar Gauss-Bonnet theory, derive its equations of motion and the principal symbol in some detail. This allows us in~\ref{sec:eff_metric} to study the characteristic equations and derive the ``effective metric" to be used as a diagnostic tool in numerical simulations. We present the numerical details and setup in section~\ref{sec:numerics}, and our numerical results in section~\ref{sec:num_results}. We also consider disformal transformations in section~\ref{sec:disf_transf}; first, we present some preliminaries and then inspect the effects of such a transformation on hyperbolicity. Finally, we conclude in~\ref{sec:conclusions}.

\section{Principal symbol and characteristics}
\label{sec:PS_char}
In this section, we briefly review some of the results in~\cite{Reall:2021voz} about the principal symbol of a scalar-tensor theory, its symmetries and the characteristics of the PDE system. In appendix~\ref{sec:hyperbolicity} we provide a pedagogical introduction to the notions of well-posedness and hyperbolicity which we hope will be useful to readers who are less familiar with these concepts.

Consider a general scalar-tensor theory of the form
\begin{align}
     S = \frac{1}{16\pi G} \int \mathrm{d}^4 x \sqrt{-g} L(g,\phi),
\end{align}
for the metric $g_{\mu \nu}$, and a scalar $\phi$ with the equations of motion given by
\begin{align}
    E^{\mu\nu} &\equiv  -\frac{16\pi G}{\sqrt{|g|}} \frac{\delta S}{\delta g_{\mu\nu}}=0, \\ 
    E &\equiv  -\frac{16\pi G}{\sqrt{|g|}} \frac{\delta S}{\delta \phi} = 0.
\end{align}
We assume the equations of motion to be second order in the metric and scalar field derivatives. Then, for an arbitrary covector $\xi_{\mu}$ the principal symbol, viewed as acting on ``polarization" vectors, is defined to be 
\begin{align}
     {\cal P}(\xi) = \left( \begin{array}{cc} P_{gg}^{\mu\nu\rho\sigma\alpha\beta}\xi_\alpha \xi_\beta & P_{gm}^{\mu\nu\alpha\beta}\xi_\alpha \xi_\beta \\ P_{mg}^{\mu\nu\alpha \beta}\xi_\alpha \xi_\beta & P^{\alpha \beta}_{mm}\xi_\alpha \xi_\beta \end{array} \right),
\end{align}
where 
\begin{align}
\label{eq:PS_defs}
    P_{gg}^{\mu\nu\rho\sigma\alpha\beta}  &\equiv \frac{\partial E^{\mu\nu}}{\partial (\partial_\alpha\partial_\beta g_{\rho \sigma})}, \qquad
    P_{gm}^{\mu\nu\alpha \beta} \equiv \frac{\partial E^{\mu\nu}}{\partial (\partial_\alpha\partial_\beta \phi)},
    \nonumber \\
    P_{mg}^{\mu \nu\alpha\beta} &\equiv \frac{\partial E}{\partial (\partial_\alpha\partial_\beta g_{\mu\nu})} ,
    \qquad 
    P_{mm}^{\alpha\beta}  \equiv \frac{\partial E}{\partial (\partial_\alpha\partial_\beta \phi)}. 
\end{align}
A more convenient notation to use is 
\begin{align}
    P_{gg}^{\mu\nu\rho\sigma}(\xi) &\equiv P_{gg}^{\mu\nu\rho\sigma\alpha\beta}\xi_{\alpha}\xi_{\beta}, \\
    P_{gm}^{\mu\nu}(\xi) &\equiv P_{gm}^{\mu\nu\alpha \beta}\xi_{\alpha}\xi_{\beta}, \\
    P_{mg}^{\mu \nu}(\xi) &\equiv P_{mg}^{\mu \nu\alpha\beta} \xi_{\alpha}\xi_{\beta}, \\
    P_{mm}(\xi)  &\equiv P_{mm}^{\alpha\beta} \xi_{\alpha}\xi_{\beta}.
\end{align}
\subsection{Symmetries of the principal symbol}
It follows by the definitions~\eqref{eq:PS_defs}, that the components of the principal symbol have the following symmetries
\begin{align}
P_{gg}^{\mu\nu\rho\sigma\alpha\beta}&=P_{gg}^{(\mu\nu)\rho\sigma\alpha\beta}=P_{gg}^{\mu\nu(\rho\sigma)\alpha\beta}=P_{gg}^{\mu\nu\rho\sigma(\alpha\beta)}, \label{eq:Psymmetry1} \\ 
P_{gm}^{\mu\nu\alpha \beta} &= P_{gm}^{(\mu\nu)\alpha \beta} = P_{gm}^{\mu\nu(\alpha \beta)}, \\
P_{mg}^{\mu \nu\alpha\beta} &= P_{mg}^{(\mu \nu)\alpha\beta}=P_{mg}^{\mu \nu(\alpha\beta)}, \\
 P_{mm}^{\alpha\beta} &= P_{mm}^{(\alpha\beta)}.
\end{align}
Furthermore, in Ref.\cite{Reall:2021voz}, the symmetries of the principal symbol that arise due to the action principle and diffeomorphism invariance were deduced. The first collection of symmetries is a consequence of the action principle, which produces the following relations 
\begin{align}
     P_{gg}^{\mu\nu\rho\sigma\alpha\beta} = P_{gg}^{\rho\sigma \mu\nu\alpha \beta}, \quad P_{gm}^{\mu\nu\alpha\beta} = P_{mg}^{\mu\nu\alpha \beta}~,
\end{align}
which implies that the principal symbol is symmetric. Moreover, the variation of the action under a compactly supported diffeomorphism generated by an arbitrary vector field yields 
\begin{align}
     P_{gg}^{\mu(\nu|\rho\sigma|\alpha \beta)}=0,  \quad \qquad P_{gm}^{\mu(\nu\alpha \beta)}=0.
\end{align}
Finally, combining these symmetries enables the writing of the principal symbol components as~\cite{Reall:2021voz}
\begin{align}
   P_{m g}^{\mu\nu\alpha\beta}
   \xi_{\alpha}\xi_{\beta}
   &=
   C^{\mu\alpha\nu\beta}
   \xi_{\alpha}\xi_{\beta}
   ,\\
   P_{gg}^{\mu\nu\rho\sigma\alpha\beta}
   \xi_{\alpha}\xi_{\beta}
   &=
   C^{\mu(\rho|\alpha\nu|\sigma)\beta}\xi_{\alpha}\xi_{\beta}
   ,
\end{align}
where $C^{\mu\nu\rho\sigma}$ has the same symmetries as the Riemann tensor
\begin{align}
\label{eq:riemann_tensor_symmetries}
   C^{\mu\nu\rho\sigma}
   =
   C^{[\mu\nu]\rho\sigma}
   =
   C^{\mu\nu[\rho\sigma]}
   =
   C^{\mu[\nu\rho\sigma]}
   ,
\end{align}
additionally, $C^{\alpha_1\alpha_2\alpha_3\beta_1\beta_2\beta_3}$ enjoys 
\begin{align}
\label{eq:antisymmetric_symmetries}
   C^{\alpha_1\alpha_2\alpha_3\beta_1\beta_2\beta_3} 
   &= C^{[\alpha_1\alpha_2\alpha_3]\beta_1\beta_2\beta_3} 
   = C^{\alpha_1\alpha_2\alpha_3[\beta_1\beta_2\beta_3]}  \nonumber \\ 
   &= C^{\beta_1\beta_2\beta_3\alpha_1\alpha_2\alpha_3}~,
\end{align}
and
\begin{align}
     C^{\alpha_1\alpha_2[\alpha_3\beta_1\beta_2\beta_3]} = C^{\alpha_1[\alpha_2\alpha_3\beta_1\beta_2\beta_3]}~,
\end{align}
as its symmetries.

These tensors are functions of $(g_{\mu\nu},\phi)$ and their first and second derivatives. Furthermore, the symmetries of $C^{\mu_1 \mu_2 \mu_3 \, \nu_1 \nu_2 \nu_3}$ imply that we can define a symmetric tensor $C_{\mu\nu}$ by 
\begin{align}
     C^{\mu_1 \mu_2 \mu_3 \, \nu_1 \nu_2 \nu_3} =-\frac{1}{2} \epsilon^{\mu_1 \mu_2 \mu_3 \rho}\epsilon^{\nu_1 \nu_2 \nu_3 \sigma} C_{\rho\sigma}.  
\end{align}
\subsection{Characteristics}
A covector $\xi_{\mu}$ is called characteristic if, and only if there exists a vector $\Omega = (\omega_{\mu \nu}, \omega)$ such that
\begin{align}
    \mathcal{P}(\xi)\Omega = 0,
    \label{eq:char}
\end{align}
expanding this equation, we have
\begin{align}
    P_{gg}^{\mu\nu\rho\sigma}(\xi) \omega_{\rho\sigma} + P_{mg}^{\mu\nu }(\xi) \omega &= 0, \label{eq:char_1} \\
    P_{mg}^{\mu\nu}(\xi)\omega_{\mu\nu} + P_{mm}(\xi)\omega &=0  \label{eq:char_2}.
\end{align}
Here, $\omega_{\mu \nu}$ is symmetric i.e., $\omega_{\mu \nu} = \omega_{\nu \mu}$. The definition of $\xi_{\mu}$ to be characteristic, as such, is insufficient in theories with gauge redundancy \cite{christodoulou2008mathematical}. In this case, due to diffeomorphism invariance a vector $\Omega = (\xi_{(\mu}X_{\nu)},0)$ solves the previous equation for all $\xi_{\mu}$ and $X_{\mu}$. This can be seen by using the symmetries of the principal symbol components~\cite{Reall:2021voz}. These unphysical modes can be ``removed" by considering only equivalence classes of solutions. Therefore, define the equivalence relation $\omega_{\mu \nu} \sim w_{\mu \nu}$ if $w_{\mu \nu} = \omega_{\mu \nu} + \xi_{(\mu}X_{\nu)}$ for some $X_{\mu}$. Hence, we can only consider the ``physical" space comprised of vectors of the form $\Omega = \left([\omega_{\mu \nu}], \omega \right)$, where $[\cdot]$ stands for equivalence class. Hence, we take $\mathcal{P}(\xi)$ to only act on such vectors \cite{christodoulou2008mathematical, Reall:2021voz}. 
With this in mind, equation~\eqref{eq:char} yields a degree six polynomial that must vanish for a characteristic $\xi_{\mu}$. That is, 
\begin{align}
    p(\xi) = (C^{-1})^{\mu \nu} \xi_{\mu}\xi_{\nu}Q(\xi) = 0~,
\end{align}
if $\xi_{\mu}$ is characteristic. The quartic polynomial $Q(\xi)$ is given by  
\begin{align}
     Q(\xi) &\equiv  \frac{C}{g} (C^{-1})^{\mu\nu} \xi_\mu \xi_\nu P_{mm}(\xi) \nonumber \\
     &+\left(2 C_{\mu\rho} C_{\nu\sigma} - C_{\mu\nu} C_{\rho\sigma} \right) P_{mg}^{\mu\nu}(\xi) P_{mg}^{\rho\sigma}(\xi),
\end{align}
where 
\begin{align}
     \frac{C}{g}&=\frac{-\sqrt{-g}}{g} \epsilon^{\mu\nu\rho\sigma} C_{0\mu} C_{1\nu} C_{2\rho} C_{3\sigma} \nonumber \\
     &= -\frac{1}{4!} \epsilon^{\mu_1 \mu_2\mu_3\mu_4} \epsilon^{\nu_1 \nu_2\nu_3\nu_4} C_{\mu_1 \nu_1} C_{\mu_2\nu_2} C_{\mu_3 \nu_3}C_{\mu_4 \nu_4}.
\end{align}
We close this section by reiterating the relation between hyperbolicity and the characteristics of the PDE system. As discussed in appendix~\ref{sec:hyperbolicity}, we can study strong hyperbolicity by finding the characteristic covector and its polarizations. \textit{Therefore, a necessary condition for well-posedness is for the polynomial $p(\xi)$ to have real roots.}

\section{The theory: scalar-Gauss-Bonnet gravity with a Ricci coupling}
\label{sec:theory}
We consider more specifically
\begin{align}
    \label{eq:action_GB}
    S=\frac{1}{16\pi}\int \mathrm{d}^{4} x \sqrt{-g}\left[R+X + f(\phi) \mathcal{G} + h(\phi)R\right],
\end{align}
where $g = \text{det}(g_{\mu \nu})$, $X=-\nabla_\mu\phi\nabla^\mu\phi/2$. We use units of $G=c=1$. The function $h(\phi)$ is a dimensionless coupling function while $f(\phi)$ is of dimension length squared. The equations of motion read  
\begin{align}
    \label{eq:metric_eqn}
    E^{\mu}{}_{\nu} \equiv & \, - T^{(\phi)\mu}{}_{\nu} + \delta^{\mu \gamma \kappa \lambda}_{\nu \alpha \rho \sigma}R^{\rho \sigma}{}_{\kappa _\lambda}\nabla_{\gamma}\nabla^{\alpha}f \nonumber \\ 
    &+(1+h) G^{\mu}{}_{\nu}+\delta^{\mu}{}_{\nu}\Box h-\nabla^{\mu}\nabla_{\nu}h = 0, \\
    
    E \equiv &\,  -\Box \phi - h^{\prime}(\phi)R - f^{\prime}(\phi)\mathcal{G} = 0,  \label{eq:phi_eqn}
\end{align} 
with 
\begin{align}
T^{(\phi)}_{\mu\nu} &= \frac{1}{2}\nabla^{\mu}\phi\nabla_{\nu}\phi-\frac{1}{4}(\nabla\phi)^2 g_{\mu\nu},
\end{align}
where $\Box \coloneqq \nabla_{\mu}\nabla^{\mu}$, $G_{\mu\nu}$ is the Einstein tensor, and $\delta^{\mu \gamma \kappa \lambda}_{\nu \alpha \rho \sigma}$ is the generalized Kronecker delta tensor. We will mainly focus on the class of theories specified by 
\begin{align}
    \label{eq:coupling_functions}
    f(\phi) =\frac{\alpha}{2}\phi^2, \quad h(\phi) = -\frac{\beta}{4}\phi^2.
\end{align}
The quadratic coupling between the scalar and $\mathcal{G}$ is the leading order contribution to the onset of the tachyonic instability that leads to black hole scalarization~\cite{Silva:2017uqg,Andreou:2019ikc,Dima:2020yac}. The coupling with $R$ has been shown to help evade binary pulsar constraints by suppressing neutron star scalarization~\cite{Ventagli:2021ubn}, to render spherical BHs radially stable~\cite{Antoniou:2021zoy,Antoniou:2022agj}, to make GR a cosmological attractor at late times~\cite{Antoniou:2020nax}, and to positively affect well-posedness~\cite{Thaalba:2023fmq,Thaalba:2024htc}. We will keep the analysis of the principal symbol and the characteristics general for any functions $f(\phi)$, and $h(\phi)$, and only impose the choice~\eqref{eq:coupling_functions} in numerical simulations. 

To compute the components of the principal symbol, we use the definitions~\eqref{eq:PS_defs}. To do so, we note that since only second-order derivatives will contribute to the symbol, we can replace covariant derivatives with partial ones. Therefore, the principal part (P.P.) of covariant derivatives acting on the scalar is trivial. For example, 
\begin{align}
    \text{P.P.}\{\nabla^{\alpha}\nabla_{\beta}h(\phi)\} = h' \partial^{\alpha}\partial_{\beta}\phi.
\end{align}
We also need to know the principal part of the Riemann, and the Einstein tensor. For the Riemann tensor, we have
\begin{align}
    \text{P.P.}\{R_{\alpha_1 \alpha_2 \beta_1 \beta_2}\} &= \frac{1}{2}\left(\partial_{\alpha_2}\partial_{\beta_1}g_{\alpha_1 \beta_2} + \partial_{\alpha_1}\partial_{\beta_2}g_{\alpha_2 \beta_1} \right. \nonumber \\
    &\left. -\partial_{\alpha_2}\partial_{\beta_2}g_{\alpha_1 \beta_1}-\partial_{\alpha_1}\partial_{\beta_1}g_{\alpha_2 \beta_2}\right),
\end{align}
while, for the Einstein tensor, recall that it can be written as
\begin{align}
   G^{\alpha}{}_{\beta} = -\frac{1}{4}\delta^{\alpha \alpha_1 \alpha_2}_{\beta \beta_1 \beta_2} R_{{\alpha_1}{\alpha_2}}{}^{{\beta_1}{\beta_2}}.
\end{align}
Now, to compute the symbol (as acting on polarization vectors), we merely need to perform the replacement 
\begin{align}
    \partial_{\alpha}\partial_{\beta}g_{\mu \nu} &\rightarrow \xi_{\alpha}\xi_{\beta}\omega_{\mu \nu}, \\
    \partial_{\alpha}\partial_{\beta}\phi &\rightarrow \xi_{\alpha}\xi_{\beta}\omega,
\end{align}
and, therefore, we find the principal symbol of the Riemann tensor to be
\begin{align}
     \left(P(\xi)\cdot \omega\right)_{\alpha_1 \alpha_2}{}^{\beta_1 \beta_2} &= \frac{1}{2}\left(\xi_{\alpha_2}\xi^{\beta_1}\omega^{\beta_2}_{\alpha_1}+\xi_{\alpha_1}\xi^{\beta_2}\omega^{\beta_1}_{\alpha_2}\right. \nonumber \\
    &\left.-\xi_{\alpha_2}\xi^{\beta_2}\omega^{\beta_1}_{\alpha_1}-\xi_{\alpha_1}\xi^{\beta_1}\omega^{\beta_2}_{\alpha_2}\right), 
\end{align}
while for the Einstein tensor, we have 
\begin{align}
    \left(P(\xi)\cdot \omega\right)^{\alpha}{}_{\beta} = \frac{1}{2}\delta^{\alpha \alpha_1 \alpha_2}_{\beta \beta_1 \beta_2}\xi_{\alpha_1}\xi^{\beta_1}\omega^{\beta_2}_{\alpha_2},
\end{align}
and finally for $\nabla^{\alpha}\nabla_{\beta}h(\phi)$ we get 
\begin{align}
    \left(P(\xi)\cdot \omega\right)^{\alpha}{}_{\beta} = h' \xi^{\alpha}\xi_{\beta}\omega,
\end{align}
thus the result follows immediately, and we have
\begin{align}
   \left(P_{gg}(\xi)\cdot \omega\right)^{\alpha}{}_{\beta}&= \frac{1}{2}(1+h)\delta^{\alpha \gamma_1 \gamma_2}_{\beta \delta_1 \delta_2}\xi_{\gamma_1}\xi^{\delta_1}\omega^{\delta_2}_{\gamma_2} \nonumber \\
   &-2\delta^{\alpha \gamma \gamma_1 \gamma_2}_{\beta \delta \delta_1 \delta_2}\xi_{\gamma_1}\xi^{\delta_1}\omega^{\delta_2}_{\gamma_2}\nabla_{\gamma}\nabla^{\delta}f, \\
    P_{mg}(\xi)^{\alpha}{}_{\beta}=P_{gm}(\xi)^{\alpha}{}_{\beta}&= \delta^{\alpha}_{\beta}h'\xi^2 - h'\xi^{\alpha}\xi_{\beta} \nonumber \\
    &+ f'\delta^{\alpha \gamma \gamma_1 \gamma_2}_{\beta \delta \delta_1 \delta_2}R_{\gamma_1\gamma_2}{}^{\delta_1\delta_2}\xi_{\gamma}\xi^{\delta},\\[2mm]
    P_{mm}(\xi) &= -\xi^2.
\end{align}

\section{Effective metric in Spherical Symmetry}
\label{sec:eff_metric}
We will focus on the characteristics of the theory~\eqref{eq:action_GB} in spherical symmetry. Hence, there are no spin-2 propagating degrees of freedom, and it is possible to reduce the characteristic equation~\eqref{eq:char} to one equation governing the scalar degree of freedom of the form
\begin{align}
    \label{eq:eff_metric}
    g_{\text{eff}}^{ab}\xi_{a}\xi_{b}\omega = 0, \quad \{a,b\} \in \{t,r\}.
\end{align}
To that end, we consider the most general spherically symmetric ansatz for the background in polar coordinates 
\begin{align}
    \mathrm{d}s^2 &= g_{tt}(t,r) \mathrm{d} t^{2}+g_{tr}(t,r) \mathrm{d}t\mathrm{d}r \nonumber \\
    &+g_{rr}(t,r) \mathrm{d} r^{2}+g_{\theta\theta}(t,r)^{2} \mathrm{~d} \Omega^{2}~,
\end{align}
and assume spherical symmetry for the scalar field as well $\phi = \phi(t,r)$.

Since we are working in spherical symmetry we can further assume that the only non-zero components of $\omega_{\mu\nu}$ are $\{\omega_{tt},\omega_{tr},\omega_{rr},\omega_{\theta\theta}\}$, and that $\omega_{\phi\phi} = \sin(\theta)^2\omega_{\theta\theta}$ with $\xi_{\mu} = (\xi_{t},\xi_{r},0,0)$. Therefore, we can diagonalize the system \eqref{eq:char_1}--\eqref{eq:char_2} by solving for the scalar degree of freedom $\omega$. We can achieve that by solving equation~\eqref{eq:char_1} for, $\{\omega_{rr},\omega_{\theta\theta}\}$, in terms of $\{\omega_{tt},\omega_{tr},\omega\}$ and substitute back into equation~\eqref{eq:char_2} to end up with an equation governing $\omega$ only from which we can read off an effective metric for the scalar degree of freedom. When $h=0$, our result matches that of~\cite{R:2022hlf}. However, the approach presented here straightforwardly generalizes to any scalar-tensor theory with second-order equations of motion. It is worth noting that a single equation for the scalar perturbation around time-independent backgrounds was obtained and used to discuss hyperbolicity in~\cite{Antoniou:2022agj}. 

As already discussed, a necessary condition for the evolution to be well-posed is for the effective metric to be Lorentzian. Thus, it constitutes a gauge-independent diagnostic tool for hyperbolicity. Therefore, in numerical considerations, we keep track of the determinant of the effective metric.

\section{Numerical Considerations} \label{sec:numerics}

We perform numerical simulations in two different gauge choices. We use the usual Schwarzschild-like coordinates as well as Painleve-Gullstrand (PG) coordinates. In this section, we describe the evolution of the different coordinate systems.  
We use the fully constrained approach and employ the method of lines to evolve in time, while the spatial derivatives are discretized using finite differences. For a detailed discussion regarding the numerical implementation and techniques used, we refer the reader to Refs.\cite{Franchini:2022ukz,Thaalba:2023fmq,Thaalba:2024htc}.

As we have discussed, the PDE system we are considering (or its reduced form as a first-order system, to be discussed below) does not always admit a (local) well-posed IVP. Computing the eigenvalues of the principal symbol is the usual method for checking if the system is hyperbolic. The system is called strongly hyperbolic if the eigenvalues are distinct and real, elliptic if any is imaginary, and parabolic if they are degenerate (for more details see~\cite{Ripley:2019irj,Ripley:2019aqj}). Here, we will be using the effective metric as a diagnostic tool to track the hyperbolic nature of the PDE system throughout its evolution. 

\subsection{Evolution equations in Schwarzschild-like coordinates}
In this case, we wish to use a less general ansatz for the metric given by
\begin{align}
    \label{eq:metric_ansatz_polar}
    \mathrm{d}s^2 = -\mathrm{e}^{2 A(t, r)} \mathrm{d} t^{2}+\mathrm{e}^{2 B(t, r)} \mathrm{d} r^{2}+r^{2} \mathrm{~d} \Omega^{2}~.
\end{align}
We introduce the following variables 
\begin{align}
    P(t,r)\equiv e^{B-A}\partial_{t}\phi, \quad Q(t,r)\equiv \partial_r \phi,
\end{align}
to reduce the equations to a first-order system that can be schematically written as 
\begin{align}
    E_{\phi} & \equiv \partial_{t} \phi - e^{A-B}P = 0, \\ 
    E_{Q} & \equiv \partial_{t} Q - \partial_{r}\left(e^{A-B}P\right) = 0,  \\ 
    E_{P} & \equiv E_{P}(\partial_{t}P; A, B, \phi, P, \partial_{r}P, Q, \partial_{r}Q), \\ 
    C_{A} & \equiv C_{A}(\partial_{r}A; B, \phi, P, \partial_{r}P, Q, \partial_{r}Q),\\  
    C_{B} & \equiv C_{B}(\partial_{r}B; B, \phi, P, \partial_{r}P, Q, \partial_{r}Q),
\end{align}
with three time-evolution equations $\{E_{\phi}, E_{Q}, E_{P}\}$, and two constraint equations $\{C_{A},C_{B}\}$.

The effective metric as a diagnostic tool, can be read from equation~\eqref{eq:eff_metric}  
\begin{align}
     \left(\alpha \xi_{t}^2 + \beta \xi_{t}\xi_{r} + \gamma \xi_{r}^2\right)\omega = 0,
\end{align}
therefore, the inverse effective metric coefficients are given by 
\begin{align}
    g_{\text{eff}}^{tt} = \alpha, \quad g_{\text{eff}}^{tr} = \frac{\beta}{2}, \quad g_{\text{eff}}^{rr} = \gamma,
\end{align}
where the explicit form of $\alpha,\beta,\gamma$ can be found in appendix~\ref{app:coeff}. 

\subsection{Evolution equations in PG-like coordinates}
Painleve-Gullstrand (PG) like coordinates in spherical symmetry are horizon penetrating, i.e., the metric functions remain regular through the formation of an apparent horizon. Therefore, they are suited to study the dynamics of BHs, and they allow the prescription of BH initial data. 
The line element is given by 
\begin{align}
    \mathrm{d} s^2&=-A(t, r)^2 \mathrm{d} t^2+\left[\mathrm{d} r+A(t, r) \zeta(t, r) \mathrm{d} t\right]^2 \nonumber \\ 
    &+r^2\left(\mathrm{d} \theta^2+\sin ^2 \theta \mathrm{d} \varphi^2\right),
\end{align}
A first-order reduction of the EOM can be achieved by introducing the following variables
\begin{align}
    Q & \equiv \partial_{r}\phi,\\
    P & \equiv \frac{1}{A} \partial_{t}\phi - \zeta Q.
\end{align}
Then, the evolution equations for $\phi$ and $Q$ can be obtained from the definitions of $P$ and $Q$ as 
\begin{align}
    E_{\phi} & \equiv \partial_{t} \phi - A(P+\zeta Q) = 0, \\ 
    E_{Q} & \equiv \partial_{t} Q - \partial_{r}\left(A(P+\zeta Q)\right) = 0,
\end{align}
and, from the EoM we obtain evolution equations for $P$ and $\zeta$, and constraint equations for $\zeta$ and $A$, which can be written schematically as 
\begin{align*}
    E_{P} & \equiv E_{P}(\partial_{t}P; A, \zeta, \phi, P, \partial_{r}P, Q, \partial_{r}Q)=0, \\
    E_{\zeta} & \equiv E_{\zeta}(\partial_{t}\zeta; \zeta, \partial_{r}\zeta, A, \partial_{r}A, \phi, P, \partial_{r}P, Q, \partial_{r}Q)=0, \\
    C_{A} & \equiv C_{A}(\partial_{r}A; \zeta, \phi, P, \partial_{r}P, Q, \partial_{r}Q)=0,\\ 
    C_{\zeta} & \equiv C_{\zeta}(\partial_{r}\zeta; \phi, P, \partial_{r}P, Q, \partial_{r}Q)=0.
\end{align*}

We can either evolve the scalar field on flat spacetime or on top of a black hole background with mass $M_{\text{BH}}$. For flat initial data, we impose
\begin{align}
    \zeta(0, r=0) = 0, \qquad 
    A(0, r=0) = 1,
\end{align}
and we impose regularity at the centre by requiring $\partial_r A|_{r=0} = 0$.
For BH initial data, we excise a region inside the apparent horizon. The latter is located at $\zeta=1$, while the location of the excision is chosen inside the horizon and it is updated to prevent the appearance of an elliptic region inside the BH. For more details, refer to the implementation of~\cite{Thaalba:2024htc}. At the excision radius, the shift $\zeta$ and the lapse $A$ are set to their GR values, namely
\begin{align}
    \zeta(0, r_{\text{exc}}) &= \sqrt{\frac{2M_{\text{BH}}}{r_{\text{exc}}}}, \\ 
    A(0, r_{\text{exc}}) &= 1.
\end{align}
In this case, we fix $M_{\text{BH}}=1$ unless otherwise stated. Finally, we do not explicitly provide the coefficients of the effective metric in these coordinates as they are rather cumbersome.

\subsection{Evolution equations in the \textit{fixing-the-equation} approach}

We also implement the \textit{fixing-the-equation}~\cite{Cayuso:2017iqc,Allwright:2018rut,Cayuso:2020lca,Franchini:2022ukz,Cayuso:2023aht,Lara:2021piy,Bezares:2021yek,Cayuso:2023aht,Gerhardinger:2022bcw,Coates:2023swo,Lara:2024rwa,Corman:2024cdr,Rubio:2024ryv} approach. The idea is that, in an effective field theory, there can be spurious degrees of freedom or spurious behaviour of actual degrees of freedom as a result of truncation at a given order in derivatives, and this can be the cause of the loss of hyperbolicity during evolution. Hence, one might be able to maintain well-posedness by ``taming" the behaviour of the specific degrees of freedom. In practice, this is done by modifying the original system [{\em e.g.}~eqs.~\eqref{eq:metric_eqn} and \eqref{eq:phi_eqn}] and supplementing it with a ``driver" equation that ``steers" the evolution away from ill-posedness and towards the true solution. Note that there is no unique choice for the driver equation.  For the system \eqref{eq:metric_eqn}-\eqref{eq:phi_eqn} with $h=0$ we consider the following ``fixed" system~\cite{Franchini:2022ukz} 
\begin{align}
	\label{eq:IS_Einstein}
	& R_{\mu \nu}-\frac{1}{2} g_{\mu \nu} R = T_{\mu \nu}^{(\phi)} + \Gamma_{\mu \nu} \,, \\
	\label{eq:IS_KG}
	& \Box\phi = \Sigma \,, \\
	\label{eq:IS_Aux}
	& \xi\Box \mbf{u} - \left( \mbf{u} - \mbf{S} \right) = 0\,,
\end{align}
where $\Gamma_{\mu \nu}$, and $\Sigma$ are auxiliary fields arranged in the vector $\mbf{u} = (\Gamma_{\mu \nu}, \Sigma)$, and $\xi$ is a constant timescales that control how the auxiliary fields approach the original theory i.e., \eqref{eq:metric_eqn}, \eqref{eq:phi_eqn}. For more details, we refer the reader to~\cite{Franchini:2022ukz}. Note that, to evaluate how the original system would behave we compute any quantity of interest using the variables evolved in the fixed system.  

\section{Numerical Results}
\label{sec:num_results}
We consider initial data (ID) that describes an approximately in-going pulse
\begin{align}
\label{eq:initialdata2_polar}
    \phi(0,r) &= a_0 \left(\frac{r}{w_{0}}\right)^2\exp\left[-\left(\frac{r-r_0}{w_0}\right)^2\right],\\
    P(0,r) &= -\frac{1}{r}\phi(0,r) - Q(0,r),
\end{align}
where $a_{0}$, $r_{0}$ and $w_0$ are constants. We will fix $r_0=25$, and $w_0=6$ unless otherwise stated.

\begin{figure}
    \centering
    \includegraphics[width=1\linewidth]{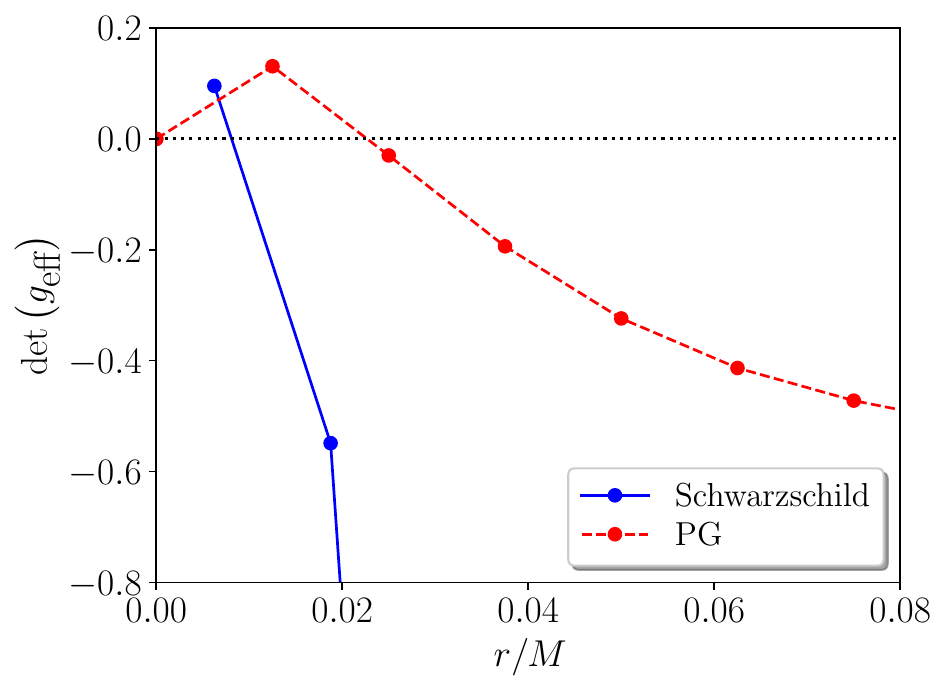}
    \caption{The plot shows the determinant of the effective metric at the time slice when hyperbolicity is lost, in different coordinates, for the same ID, and coupling constants with $\alpha/M=0.25$, and $\beta=0$. We observe that the sign of the determinant changes at the same location in both coordinates within a tolerance of grid spacing.}
    \label{fig:g_eff_sign}
\end{figure}

We start with a small pulse on flat spacetime, with particular values of couplings, $\alpha/M=0.25$, and $\beta=0$, which are known to lead to loss of hyperbolicity during this evolution of the data above. We evolve the data using the two different coordinate systems and verify that the effective metric does indeed change sign when hyperbolicity is lost. Moreover, as expected, the change of sign happens at the same location for both coordinate systems.  As depicted in Fig.~\ref{fig:g_eff_sign},  the determinant changes sign at the same radius (within grid spacing) when the system becomes elliptic and when it switches back again (from positive to negative). This is a demonstration that the gauge-independent property of the determinant is manifest in numerical simulations. We have performed the same runs with double and quadruple resolutions, in order to check that the sign change happens at the same radius up to the grid spacing of the two different codes,  and we have found consistency in the results. It is worth noting that a small discrepancy in the localization of the radius might also arise due to the different definitions of the time coordinate in Schwarzschild and PG coordinates.

Next, we evolve the ID exploring the parameter space of the theory. We keep track of the effective metric to probe the loss of hyperbolicity and check whether or not it can be attributed to the gauge choice. First, in Schwarzschild coordinates, we reexamine the parameter space already explored in~\cite{Thaalba:2023fmq}. We find that the effective metric criterion reproduces the results found there, and hence the observed behaviour is not due to the gauge choice. This might indicate that the ill-posedness observed in the literature in various studies of scalar Gauss-Bonnet gravity~\cite{Ripley:2019hxt,Ripley:2019irj,Ripley:2020vpk,East:2020hgw,East:2021bqk,Corman:2022xqg,AresteSalo:2022hua,R:2022hlf,Corman:2024vlk,Corman:2024cdr,Lara:2024rwa} is due to the physical, rather than the gauge modes present in the theory. Furthermore, this provides further evidence that additional interactions that would be expected to be present in an EFT, such as the Ricci coupling, can be crucial for well-posedness.

\begin{figure*}[t]
     \centering
    \includegraphics[width=.47\linewidth]{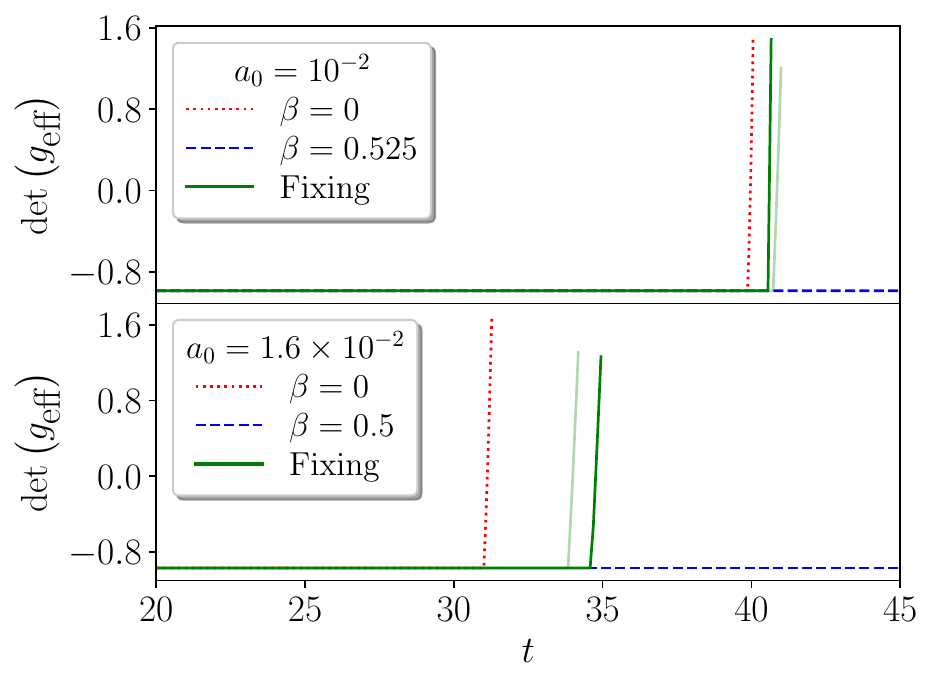}
    \hspace{5mm}
    \includegraphics[width=.47\linewidth]{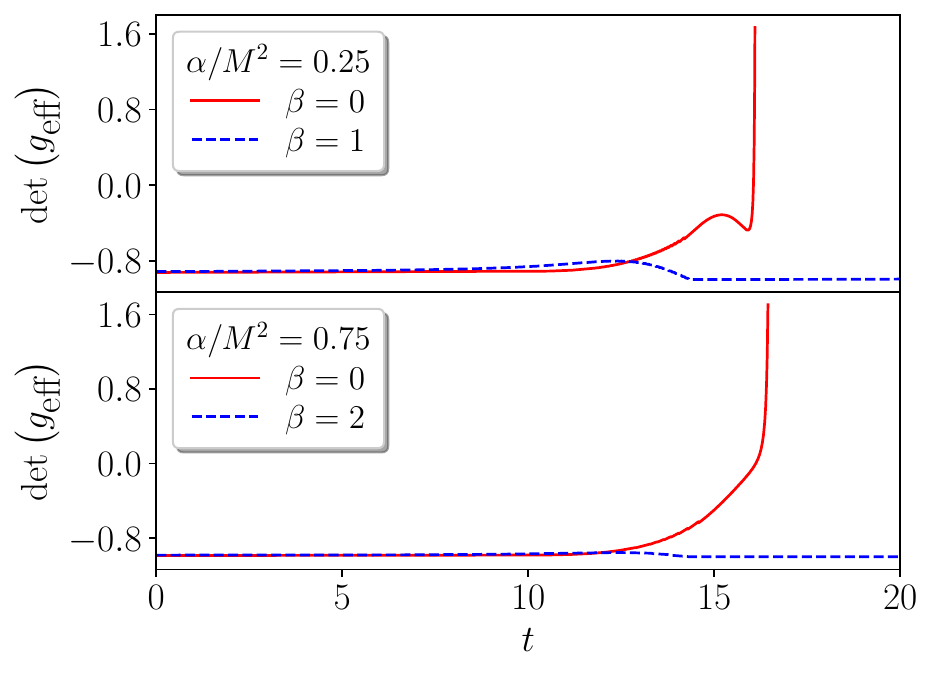}
    \caption{Both plots show the maximum of the determinant of the effective metric $\text{det}\left(g^{ab}_{\text{eff}}\right)$ in space for different values of $\beta$. \textit{Left}: The evolution was performed in polar coordinates. The choice of coupling constant is $\alpha/M^2=0.25$. The top panel is for $a_0 = 10^{-2}$ for which the evolution ceases to be well-posed in the $\beta=0$ case, as the determinant of the effective metric grows and crosses zero. However, the effective metric remains Lorentzian for $\beta=0.525$, and the evolution remains hyperbolic. We observe similar behaviour in the bottom panel. In this case, the choice of the ID ($a_0 = 1.6 \times 10^{-2}$) leads to the formation of an apparent horizon with a negligible scalar field. We also show the effective metric in the fixed theory for $\beta = 0$ for two different timescales that increase for darker shades of green. \textit{Right}: The evolution was performed in PG coordinates with BH ID and a scalar field perturbation. We observe a similar trend as the Schwarzschild case. The system is no longer hyperbolic when the effective metric stops being Lorentzian. The top panel is for ID with $a_0 = 5\times 10^{-3}$, $\alpha/M^2=0.25$. The final state (when the evolution remains well-posed) is a Schwarzschild BH. We have the same ID in the bottom panel, with $\alpha/M^2=0.75$ but the final state is a scalarized BH.}
    \label{fig:g_eff}
\end{figure*}

In the left panel of Fig.~\ref{fig:g_eff}, we present the characteristic trend of the determinant of the effective metric in Schwarzschild coordinates. Here, we consider $\alpha/M^2=0.25$ and different values of $\beta=\{0, 0.5, 0.525\}$ to display the signature of the effective metric in distinct scenarios. In the top plot, the amplitude of the ID is $a_0=10^{-2}.$ When hyperbolicity is lost (with $\beta=0$), we notice the growth of the determinant and its value crossing zero and becoming positive, i.e., the effective metric of the system is no longer Lorentzian. On the other hand, when $\beta=0.525$, the effective metric remains Lorentzian, and the final state of the evolution is flat spacetime. In the bottom panel, we consider $a_0=1.6\times10^{-2}$, which collapses into a BH with negligible scalar field when $\beta=0.5$; otherwise ($\beta=0$), the evolution is ill-posed, as indicated by the effective metric.

\begin{figure}
    \centering
    \includegraphics[width=1\linewidth]{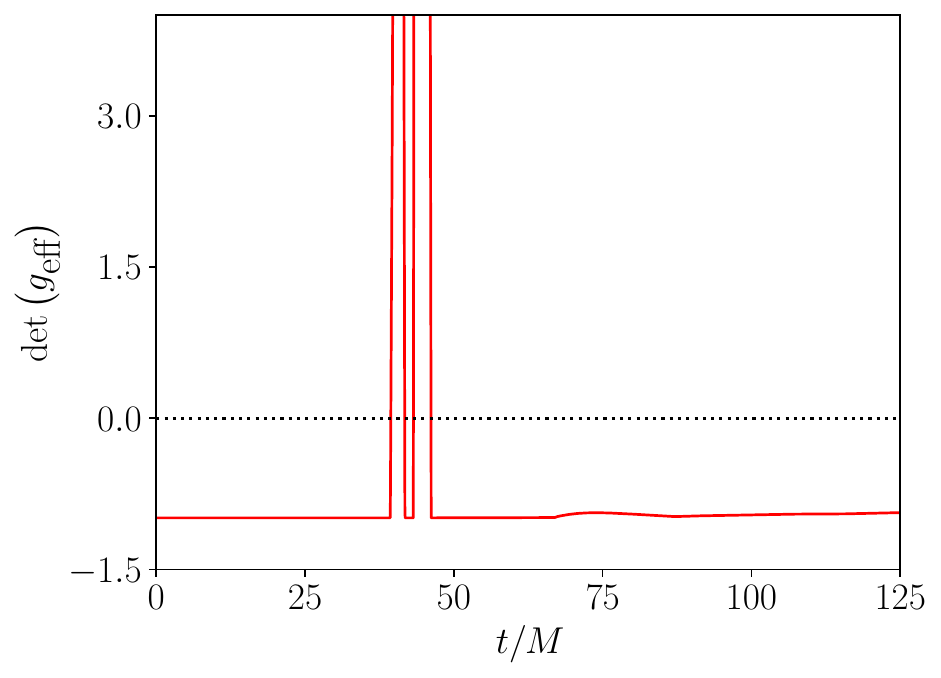}
    \caption{We plot the behaviour of the maximum of the determinant of the effective metric reconstructed with the fixed theory right for ID with $a_0=9.5\times 10^{-3}$, and $\alpha/M^2 = 0.25$, $\beta=0$. We see that at some point the effective metric flips sign and becomes non-Lorentzian, but as we are using the fixing procedure we can follow the evolution which results in flat spacetime with the effective metric behaving nicely.}
    \label{fig:geff_fixed_flat}
\end{figure}

To make sure that the behaviour of the effective metric near the loss of hyperbolicity is not affected by the loss of convergence and accumulating errors as the system changes character from hyperbolic to elliptic, we also compute the same determinant in simulations of the same ID using the fixing-the-equations approach. In this case, the evolution of the ``fixed'' system remains hyperbolic, but the determinant tracks the hyperbolicity properties of the original system.
The behaviour of the effective metric evaluated using the ``fixed" system is depicted in the left panel of Fig.~\ref{fig:g_eff} for different values of the time scale $\xi$. We see the same trend produced by the ``original" system indicating that the ill-posedness is not caused by the coordinate system or the numerical instability caused by the system changing character. Moreover, in Fig.~\ref{fig:geff_fixed_flat}, we examine the trend of the determinant in the fixed system for which the end state of the evolution is flat spacetime. We observe that the original system becomes elliptic at some point, as suggested by the effective metric. However, as expected, as the system evolves to flat geometry the effective metric settles down and becomes Lorentzian again. 

We also study BH ID using PG coordinates. Two examples are given in the right panel of Fig.~\ref{fig:g_eff}, where we have $\alpha/M^2=0.25$ (top panel), and $\alpha/M^2=0.75$ (bottom panel). The top and bottom panels have $a_0=5\times10^{-3}$. In both cases when $\beta=0$ we observe similar proclivity for the effective metric to become non-Lorentzian, but in the occasions where the evolution is hyperbolic ($\beta>0$) the end states are either a Schwarzschild (top panel) or a scalarized (bottom panel) BH. 

\begin{figure}
    \centering
    \includegraphics[width=1\linewidth]{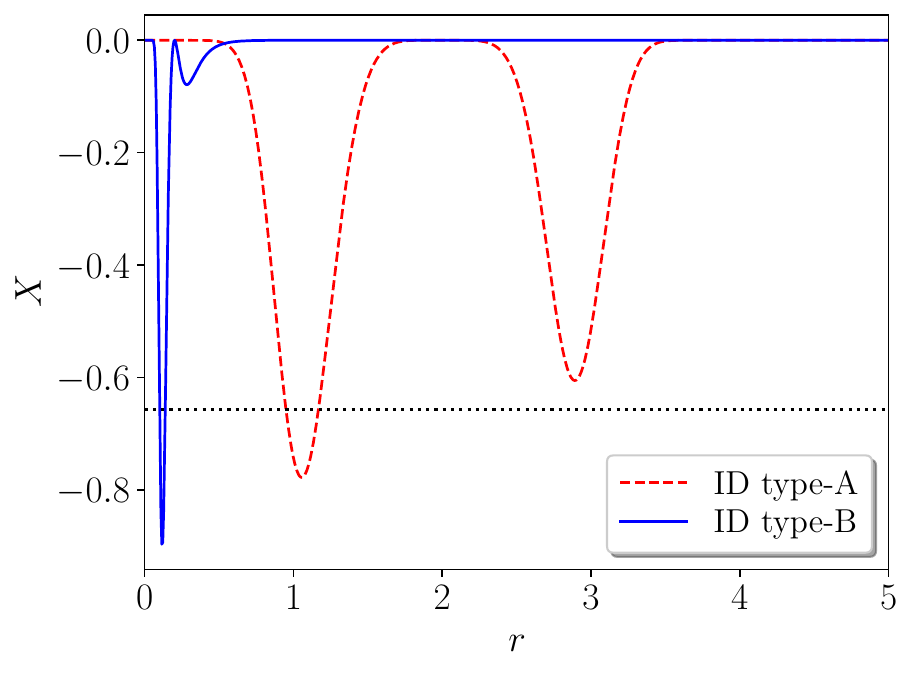}
    \caption{The plot shows the kinetic term at $t=0$. We have type-A ID with $a_0=0.85$, $r_0=2$, and $w=1$. We observe that the effective metric is not Lorentzian, as $X$ is less than the threshold for some $r$, indicating that the problem is ill-posed after performing the disformal transformation. Nevertheless, the original formulation i.e., GR with a minimal scalar produces a well-posed evolution. Similar behaviour is observed for $a_0$ approximately in $[0.78,1.98]$ without being hidden by an apparent horizon. For type-B ID we have $a_0=5\times 10^{-3}$, $r_0=2$, and $w=1$. For $a_0$ approximately in $[5\times 10^{-3},0.7]$, the evolution is akin to the case shown without forming an apparent horizon.}
    \label{fig:kinetic_term}
\end{figure}

\section{Disformal transformations}
\label{sec:disf_transf}

The key advantage of using the determinant of the effective metric, or more generally the principal symbol analysis of Ref.~\cite{Reall:2021voz}, is to assess well-posedness in a gauge invariant manner. This makes it a more trustworthy diagnostic than studying the evolution in a particular gauge. Nonetheless, it cannot be seen as a definitive criterion of whether or not a theory is well-posed --- after all well-posedness is a statement about a particular initial value formulation of a theory. One can perform field redefinitions that affect the formulation of the IVP and its hyperbolicity. 
An invertible linear transformation of the PDE system should leave the hyperbolic nature of the system intact. If the transformation is derivative-dependent however, then, mostly, this is no longer the case.
To illustrate this point with an example, we consider here the effect of disformal transformations on hyperbolicity.

\subsection{Preliminaries}
Disformal transformations were initially introduced by Bekenstein in~\cite{Bekenstein:1992pj}. Such transformations are defined through a field redefinition of the form 
\begin{align}
    \label{eq:disf_transf}
    g_{\mu \nu} \rightarrow \mathcal{A}(\phi, X)g_{\mu \nu} + \mathcal{B}(\phi, X)\nabla_{\mu}\phi\nabla_{\nu}\phi \eqqcolon \tilde{g}_{\mu \nu},
\end{align}
with disformal functions $\mathcal{A}$, and $\mathcal{B}$. In the case of $\mathcal{A} = \mathcal{A}(\phi)$, and $\mathcal{B}=0$, we retrieve the usual conformal transformations. The disformal functions now depend implicitly on the metric and how the scalar field changes in spacetime through the kinetic term, not only on the field itself. 

The disformal field redefinitions should satisfy some ``reasonable" conditions to produce a ``well-behaved" metric.  We impose that the disformal metric is invertible with a non-singular volume element, given by 
\begin{align}
    g^{\mu \nu} &\rightarrow \frac{1}{\mathcal{A}}\left(g^{\mu \nu} - \frac{\mathcal{B}}{\mathcal{A}-2\mathcal{B}X}\nabla^{\mu}\phi\nabla^{\nu}\phi\right), \\ 
    \sqrt{-g} &\rightarrow \mathcal{A}^2\left(1-2\mathcal{B}X/\mathcal{A}\right)^{1/2}\sqrt{-g},
\end{align}
from which, for positive $\mathcal{A}$ (the case we will consider), we have the condition
\begin{align}
    \mathcal{A}-2\mathcal{B}X > 0.
\end{align}
Two different representations of a Lagrangian related by a disformal transformation will have the same degrees of freedom and physical solutions for invertible transformations~\cite{Takahashi:2017zgr}. An additional constraint on the functions $\mathcal{A}$, and $\mathcal{B}$ arises as a consequence of invertibility, given by~\cite{Zumalacarregui:2013pma} 
\begin{align}
    \label{eq:invertible}
    \mathcal{A}\left(\mathcal{A}-X\partial_X \mathcal{A} + 2 X^2 \partial_X \mathcal{B}\right) \neq 0.
\end{align}
Note that if the functions $\mathcal{A}$, and $\mathcal{B}$ are independent of the kinetic term $X$, then the relation between $g_{\mu \nu}$, and $\tilde{g}_{\mu \nu}$ can be trivially inverted  
\begin{align}
    g_{\mu \nu} = \frac{\tilde{g}_{\mu \nu} - \mathcal{B}(\phi)\nabla_{\mu}\phi\nabla_{\nu}\phi}{\mathcal{A}(\phi)},
\end{align}
therefore, the transformation is invertible for a non-zero $\mathcal{A}$.
\subsection{Disformal transformations and hyperbolicity}
To examine the effect of field redefinitions on hyperbolicity we consider GR minimally coupled to a scalar field (i.e., $f=h=0$ in the action \eqref{eq:action_GB}), and perform a transformation of the form~\eqref{eq:disf_transf} such that $\mathcal{A} = a$, and $\mathcal{B} = b$ for some constants $a,b$. The resultant theory is~\cite{Bettoni:2013diz} 
\begin{align}
\label{eq:disf_action}
    S&=\frac{1}{16\pi}\int \mathrm{d}^{4} x \sqrt{-g}\left[\mathcal{G}_2 + \mathcal{G}_4 R \right. \nonumber \\
    &\left.+\partial_X \mathcal{G}_4 \left(\left(\Box \phi\right)^2 - \left(\nabla_{\mu}\nabla_{\nu}\phi\right)^2\right)\right],
\end{align}
where,
\begin{align}
    \mathcal{G}_{2} &= \frac{X}{a \mathcal{S}}, \\ 
    \mathcal{G}_{4} &= a\mathcal{S}, \quad \mathcal{S} = \sqrt{1-\frac{2 b X}{a}}.
\end{align}
Hence, we end up with a Horndeski theory with a specific choice of the functions $\mathcal{G}_2$, and $\mathcal{G}_4$. For this example, we have chosen the disformal functions to be constants for simplicity. The considered choice is rigid and might lead to some of the conditions we discussed being dynamically violated in an evolution.
Nonetheless, in what follows, we will analyse the behaviour of this theory for some ID at $t=0$ and ensure that these conditions are fulfilled.

It was shown in~\cite{Papallo:2017qvl} that a theory with $\partial_{X}\mathcal{G}_{4}\neq0$ is not strongly hyperbolic in the generalized harmonic gauge. To understand the effect of field redefinitions in a gauge-independent manner, we proceed as in the case of scalar Gauss-Bonnet and compute the effective metric in this theory. The equations of motion and the principal symbol are given in appendix~\ref{app:disf_eqns_PS}.

Now, we can analyse the characteristic equation as we have done in the case of scalar Gauss-Bonnet, and find the determinant of the inverse effective metric (in Schwarzschild-like coordinates~\eqref{eq:metric_ansatz_polar}), which yields
\begin{equation}
\begin{split}
    \text{det}\left(g^{ab}_{\text{eff}}\right) &= \frac{e^{-2 (A+B)}}{(a-2 b X)^4}\left(-a^2+ a^2 b^2 X^2 \right. \\
    &\left.-2 a b^2 X^2 \mathcal{S}-4 a b^3 X^3+4 b^4 X^4\right),
\end{split}
\end{equation}
and the kinetic term is given by
\begin{align}
    X = \frac{1}{2} \left(e^{-2 A} \left(\frac{\partial \phi }{\partial t}\right)^2-e^{-2 B}
   \left(\frac{\partial \phi }{\partial r}\right)^2\right).
\end{align}
For concreteness, we fix $a=b=1$, and find the values of $X$ for which the effective metric ceases to be Lorentzian, 
\begin{align}
    \label{eq:disf_transf_cond}
    \text{det}\left(g^{ab}_{\text{eff}}\right) \geq 0 \iff X \lessapprox -0.6573.
\end{align}
To study the effect of disformal transformations on hyperbolicity, we evolve GR in polar coordinates using two types of ID that lead to condition~\eqref{eq:disf_transf_cond} being satisfied initially, whilst maintaining the invertibility of the transformation. That is, the physics is left invariant under the disformal change of variables. If we take the ID for the scalar field to be
\begin{itemize}
    \item Type-A: 
    \begin{align}
            \phi(0,r) &= a_0 \exp\left[-\left(\frac{r-r_0}{w_0}\right)^4\right], \quad P(0,r) = 0,
        \end{align}
    \item Type-B:
        \begin{align}
             \phi(0,r) &= \frac{a_0}{r^5}\left(\exp(1/r)-1\right)^{-1}, \quad P(0,r) = 0.        
        \end{align}
\end{itemize}
Then the effective metric will change signature at some $r>0$, as the value of $X$ is below the threshold as shown in Fig.~\ref{fig:kinetic_term}. Note that the invertibility condition~\eqref{eq:invertible} is trivially satisfied. Therefore, for such ID, the problem will be ill-posed, while in the original formulation (GR with a minimal scalar), the theory is always well-posed.

\section{Conclusions}
\label{sec:conclusions}

We have studied numerically the well-posedness of spherical dynamics in scalar Gauss-Bonnet gravity with additional Ricci coupling, using the gauge-invariant method of Ref.~\cite{Reall:2021voz}. In this setup loss of hyperbolicity can be probed by calculating when an effective metric becomes degenerate. 

We have used flat and BH ID for various coupling constants and explored the same part of the parameter space as in Refs.~\cite{Thaalba:2023fmq,Thaalba:2024htc}, performing simulations in both of the gauge choices used therein. We find that the determinant of the effective metric changes sign in cases for which loss of hyperbolicity was reported in Refs.~\cite{Thaalba:2023fmq,Thaalba:2024htc}, while it remains nondegenerate for sufficiently large values of the Ricci coupling. This indicates that it is the behaviour of the physical degrees of freedom, rather than a gauge choice, that is responsible for dynamically changing the character of the equations from hyperbolic to elliptic. Our results provide further evidence that the Ricci coupling has a positive effect on hyperbolicity in spherical symmetry and that, more broadly, including additional couplings that would be present in an EFT can be crucial for having a well-posed IVP beyond GR.

Although the (sign of the) determinant of the effective metric is a gauge invariant probe of well-posedness, one could have concerns about accuracy when calculating it numerically in a specific gauge near the loss of hyperbolicity. To address this
we have performed additional simulations utilising the fixing-the-equations approach, where evolution remains hyperbolic, and monitored the evolution of the effective metric of the original system. We found that the determinant changes sign in the same fashion as it did in the original theory. This indicates that our earlier results are trustworthy and that the growth of the determinant was not due to the accumulating numerical error close to the elliptic region.  
We have also considered initial data for which one expects that the endpoint is flat space, but for which the original system of equations becomes elliptic during evolution. We have then verified that in the fixed system evolution proceeds without problems and flat space is indeed the endpoint. We computed the effective metric of the original system using the fixed system and found that it becomes non-Lorentzian when the system changes character but then goes back to being Lorentzian as the evolution proceeds in the fixed system for longer times. 

Additionally, we highlighted a limitation in using the determinant of the effective metric as a probe of hyperbolicity. Although it is gauge invariant it is not invariant under field redefinitions, which can indeed affect the character of the evolution equations. As an illustrative example, we analysed the effect of disformal transformations on a minimally coupled scalar field to GR. This transformation maps GR to a certain Horndeski theory which is not always well-posed. We evolved GR for certain choices of the ID and ensured that the transformation is invertible initially to have the same physics in both theory representations. We found that the effective metric of the Horndeski theory indicates that this ID will lead to elliptic equations, and hence, an ill-posed problem.

It would be fruitful to generalize this analysis to $3+1$ dimensions. This will provide a useful diagnostic tool in the case of $3+1$ numerical codes. Nonetheless, the $3+1$ case is more complicated and generally it would not be possible to reduce the system to a scalar equation with an effective metric as the scalar and gravitational degrees of freedom will mix beyond spherical symmetry. Alternatively, one would need to find the roots of the quartic polynomial and check that they are always real.

\begin{acknowledgments}
We would like to thank  \'Aron D. Kov\'acs for collaborating in the early stages of this work. 
TS acknowledge partial support from the STFC Consolidated Grant nos. ST/V005596/1 and ST/X000672/1. MB acknowledge partial support from the STFC Consolidated Grant no. ST/Z000424/1.
\end{acknowledgments}

\appendix
\section{Hyperbolicity}
\label{sec:hyperbolicity}
In what follows, we review the notion of well-posedness, which we mostly base on \cite{Kreiss, Sarbach:2012pr}. In general, Hadamard criterion \cite{hadamard} states that any PDE system is well-posed if it possesses a unique solution that depends continuously on the initial data, i.e., any ``small changes" in the initial data only lead to ``small changes" in the solution. We consider the following approach to establish and examine the well-posedness of an IVP. We translate the conditions of uniqueness and continuous dependence on the initial data to algebraic conditions on the principal symbol of the system (i.e., the highest derivative terms). We will mainly focus on first and second-order PDEs that describe an initial value problem. We will define the concepts of weak and strong hyperbolicity and the necessary algebraic conditions that lead to a well-posed IVP.

\subsection{First order system}
Consider an IVP in $d+1$ spacetime dimensions of the following form 
\begin{equation}
\begin{aligned}
    P^{\mu}\partial_{\mu}u(t,x) + Q u(t,x)  &= 0, \quad (t,x) \in \mathbb{R}^{d+1}~, \label{eq:CP1} \\ 
    u(0,x) &= f(x),  \quad x \in \mathbb{R}^{d}~,
\end{aligned}
\end{equation}
where $u$ is an $n$--dimensional column vector of all the variables in the system. $P^{\mu}$ and $Q$ are real constant $n \times n$ matrices. A formal solution of the system~\eqref{eq:CP1}
can be obtained by performing a Fourier transform in space
\begin{align}
    \hat{u}(t, \xi) = \frac{1}{(2\pi)^{d/2}} \int_{\mathbb{R}^d} e^{-i\xi \cdot x} u(x,t) \mathrm{d}x, \quad \omega \in \mathbb{R}^d, 
\end{align}
then, the IVP~\eqref{eq:CP1} becomes 
\begin{equation}
\begin{aligned}
   \partial_{t}\hat{u} - i A (\xi_i)\hat{u} &= 0,\\ 
    \hat{u}(0,\xi) &= \hat{f}(0,\xi), 
\end{aligned}
\end{equation}
where $A(\xi_i) = (P^0)^{-1}(-P^i\xi_i + iQ),$ and we have assumed that $P^0$ is invertible. This IVP has a solution of the form 
\begin{align}
    \hat{u}(t,\xi_i) = e^{iA(\xi_i)t}\hat{f}(\xi_i),
\end{align}
which we can inverse Fourier transform to obtain a solution of the original problem~\eqref{eq:CP1}
\begin{align}
    u(t,x) = \frac{1}{(2\pi)^{d/2}}\int_{\mathbb{R}^d} e^{i\xi \cdot x}e^{iA(\xi_{i})t}\hat{f}(\xi_i)\mathrm{d}\xi.
    \label{eq:formal_sol}
\end{align}
This solution is formal, in that we can not guarantee its convergence. Even if the initial data $f$ is chosen such that $\hat{f}$ is smooth and decays to zero as $\vert\xi\vert \coloneqq \sqrt{\xi^i \xi_i} \rightarrow \infty$, the term $\abs{e^{iA(\xi_{i})t}}$ might diverge in the limit $\vert\xi\vert \rightarrow \infty$. On the other hand, if we require that $\abs{e^{iA(\xi_{i})t}}$ is bounded by an exponential in time that is independent of $\xi_i$ as  
\begin{align}
   \abs{e^{iA(\xi_{i})t}} \leq c e^{\kappa t}, \quad \forall \xi_i, \forall t \ge 0, \label{eq:boundedness}
\end{align}
for some constants $c \geq 1$, and $\kappa \in \mathbb{R}$, the integral will converge. This bound will imply, by Parseval's identity, the following 
\begin{align}
    \abs{u} = \abs{\hat{u}} = \abs{e^{iA(\xi_{i})t}\hat{f}(\xi_i)} \leq c e^{\kappa t} \abs{f},
\end{align}
and if the initial data is $L^2$, \footnote{A function $f$ is an $L^2$ function if $x \rightarrow \vert f(x) \vert ^2$ is Lebesgue-integrable over $\mathbb{R}^d$.} then 
\begin{align}
    \normL{u} \leq c e^{\kappa t} \normL{f}.
\end{align}
Hence, the norm of the initial data controls that of the solution, precisely the condition for well-posedness.

To arrive at the algebraic conditions for well-posedness, let us consider the behaviour of the solution at high frequencies, i.e., in the limit $\vert\xi\vert\rightarrow \infty$, and let $t = \tilde{t}/\vert\xi\vert$, then equation~\eqref{eq:boundedness} becomes
\begin{align}
    \abs{e^{iB(\tilde{\xi}_{i})\tilde{t}}} \leq c, \quad \tilde{\xi}_{i} = \frac{\xi_i}{\vert\xi\vert}, 
\end{align}
where
\begin{align}
    B(\xi_i) = -(P^0)^{-1}P^i\xi_i
\end{align} 
corresponds to the highest derivative terms. 

Now, let $b$ be an eigenvector of the matrix $B(\xi_i)$ with eigenvalue $\lambda$, then
\begin{align}
    e^{iB(\tilde{\xi}_{i})\tilde{t}} b = e^{i\lambda\tilde{t}} b,
\end{align}
since the $B$ matrix is real then $\lambda ^*$ is an eigenvalue as well with the eigenvector $b ^ *$, therefore we also have 
\begin{align}
    e^{iB(\tilde{\xi}_{i})\tilde{t}} b = e^{i\lambda ^*\tilde{t}} b,
\end{align}
and to prevent exponential growth we must have $\text{Im}(\lambda) = 0$. The concept of weak hyperbolicity is based on this condition, that is to say, the IVP~\eqref{eq:CP1} is weakly hyperbolic if, and only if, the matrix $B$ has only real eigenvalues for all unit-norm $\xi_i$. Nonetheless, this requirement is insufficient. For assume that the matrix $B$ is not diagonalizable, i.e., there exists an $m \times m$ (for some $2 \leq m \leq n$) Jordan block in its Jordan normal form, then $\abs{e^{iB(\tilde{\xi}_{i})\tilde{t}}}$ may exhibit polynomial growth of the form $\abs{\xi}^m$, see Ref.~\cite{Kreiss}. Therefore, to establish well-posedness, the matrix $B$ needs to be diagonalizable with real eigenvalues~\cite{Kreiss}. Consequently, we will state that a PDE system is said to be strongly hyperbolic if, and only if, its matrix $B$ is diagonalizable with real eigenvalues. 

Hyperbolicity is related to the existence of characteristics. Let $\xi_0$ be an eigenvalue of $B(\xi_i)$ with an eigenvector $\omega$. Then this is equivalent to having 
\begin{align}
    \mathcal{P(\xi)}\omega \equiv P^{\mu}\xi_{\mu}\omega = 0, \label{eq:Pxi}
\end{align}
and this equation has a non-trivial solution if, and only if, $\text{det}\mathcal{P}(\xi) = 0$. That is, a non-trivial solution exists if, and only if, $\xi$ is a characteristic covector.

\subsection{Second order system}
We consider a linear second-order system in $d+1$ spacetime dimensions of the form 
\begin{align}
      P^{\mu \nu} \partial_{\mu} \partial_{\nu} u + Q^{\mu}\partial_{\mu}u +Ru = 0,
\end{align}
as in the first-order system case, $u$ is an $n$-dimensional vector, with $P^{\mu \nu}$, $Q^{\mu}$, and $R$ are $n \times n$ real constant matrices. The principal symbol is defined as 
\begin{align}
    \mathcal{P}(\xi) \equiv P^{\mu \nu} \xi_{\mu} \xi_{\nu}.
\end{align}

We will only outline the analysis for this case (for more details, see \cite{Kreiss, Sarbach:2012pr, Papallo:2017qvl}). We initially proceed as in the first-order case by performing a spatial Fourier transform and then writing the resultant equation in a first-order form as \cite{Papallo:2017qvl}
\begin{align}
    \hat{w} = i A(\xi_i)\hat{w}, \quad \hat{w}^T = \left(\sqrt{1+\abs{\xi}^2}\hat{u}, -i \hat{u}_t\right),
\end{align}

for some $2n \times 2n$ matrix $A(\xi_i)$. Therefore, we can now establish a formal solution. After which, we can study the behaviour of high-frequency solutions to identify the dominant part of $A(\xi)$ in the $\abs{\xi} \rightarrow \infty$ limit, which yields
\begin{align}
    B(\xi_i) = \begin{pmatrix} 0&I \\ -(P^{00})^{-1}P^{ij}\xi_{i}\xi_{j}&-2(P^{00})^{-1}P^{0i}\xi_{i} \end{pmatrix}.
\end{align}
The same definitions for weak and strong hyperbolicity carry over. To connect our discussion of hyperbolicity in second-order systems to the characteristics of such PDEs, we take $\xi_{0}$ to be an eigenvalue of $B$ with an eigenvector $v = (\omega, \omega')$. The eigenvalue equation will then yield, 
\begin{align}
    P^{\mu \nu} \xi_{\mu} \xi_{\nu} \omega = 0 , \quad \omega ' = \xi_{0}\omega, 
\end{align}
which has a non-trivial solution if, and only if, $\xi_{\mu}$ is characteristic. So, we can demonstrate strong hyperbolicity by finding the characteristic covectors and their ``polarization" $\omega$. 

\subsection{Variable coefficients and non-linear PDEs}

Up to this point, we have only discussed linear PDE systems with constant coefficients. An obvious question is how to generalize the previous discussion to non-linear systems with variable coefficients. The ``localization principle" addresses PDE systems with variable coefficients. The principle of localization is based on well-posedness being a high-frequency question, i.e.~on the premise that we should only worry about the high-frequency (very short wavelength) solutions. If we assume that the coefficients are smooth, then in this regime (of high frequencies), we can ``localize" the coefficients of the PDE at a point $p$, and consider them ``frozen" to their value at $p$. Then the localization principle states that ``A necessary condition for local well-posedness of the varying coefficient equations near $p$ is that the frozen coefficient equation should be locally well-posed for all points $p$ (with additional smoothness requirements on the eigenvalues and vectors)." \cite{Kreiss, Sarbach:2012pr}.

On the other hand, non-linear problems introduce, in general, many complications. For instance, non-linearities may cause the solution to blow up in finite time or lead to the crossing of the characteristics producing shocks \cite{Kreiss, Sarbach:2012pr}. Consequently, it is almost always only possible to establish the existence of a solution for some time interval. If a non-linear system is well-posed, then a small perturbation $u_{1}$ of the solution $u_0$ as $u = u_0 + \epsilon u_1$ will produce a well-posed linear system for $u_1$. Hence, a necessary condition for the well-posedness of the non-linear system is that when linearizing around any solution, the consequent linear problem is well-posed. More importantly, the converse is true \cite{Kreiss, Sarbach:2012pr} for a quasilinear problem. This is known as the linearization principle: ``The quasilinear problem is well-posed in the neighbourhood of a solution if all the linearized problems around that solution are well-posed". As we are interested in completely nonlinear PDEs it is worth pointing out that such systems are more complicated to get a handle on and require more advanced machinery to tackle. The treatment of such problems is available in~\cite{taylor1991paradifferential} (see e.g., chapter $5$). The main difference between quasilinear and completely nonlinear systems lies in the required order of regularity imposed on the initial data. That subtlety aside, strong hyperbolicity as defined above still guarantees (local) well-posedness.

\section{Effective metric coefficients}
\label{app:coeff}
Here, we provide the coefficient $\alpha$, $\beta$, and $\gamma$ of the effective metric in Schwarzschild-like coordinates
\begin{widetext}
\begin{align}
    \alpha &= \frac{1}{r^2 e^{2 A+2 B} \left(e^{2 B} (h+1) r-4 Q f'\right)^2} \left( e^{6 B} (h+1) r^2 \left(12 f' h'+r^2 \left(3 \left(h'\right)^2+h\right)+r^2\right)
    +384 Q \left(f'\right)^3 \frac{\partial B}{\partial r} \right. \nonumber \\
    &\left.-e^{4 B} r \left(f' \left(P^2 \left(4 r f'+r^3 h'\right)-12 (h+1) \left(8 f' \frac{\partial B}{\partial r}+r h' \left(2 r \frac{\partial B}{\partial r}-1\right)\right)\right)
    \right. \right. \nonumber \\
    &\left.\left.+8 Q f' \left(12 f'h'+r^2 \left(3 \left(h'\right)^2+h\right)+r^2\right)+Q^2 r \left(4 f'+r^2 h'\right) \left(f' \left(4 h''+1\right)-4 f'' h'\right)\right) \right. \nonumber \\
   &\left.+4 e^{2 B} f' \left(-24 f'
   \frac{\partial B}{\partial r} \left(Q \left(4 f'+r^2 h'\right)+h r+r\right)+P^2 r^2 f'+Q r \left(24 f' h'+Q r \left(f' \left(4 h''+5\right)-4 f''
   h'\right)\right)\right) \right) \\  \nonumber \\
   
   \beta &= \frac{4 e^{-2 (A+B)} \left(4 \left(e^{2 B}-1\right) f'+e^{2 B} r^2 h'\right) \left(P Q r^2 e^{A+B} \left(f' \left(2 h''+1\right)-2 f'' h'\right)-12 f' \frac{\partial
   B}{\partial t} \left(e^{2 B} (h+1) r-4 Q f'\right)\right)}{r^2 \left(e^{2 B} (h+1) r-4 Q f'\right)^2} \\ \nonumber \\ 
   
   \gamma &= \frac{1}{e^{4 B} r^2 \left(e^{2 B} (h+1) r-4 Q f'\right)^2} \left(e^{4 B} r \left(12 (h+1) f' \left(8 f' \frac{\partial A}{\partial r}+r h' \left(2 r \frac{\partial A}{\partial r}+1\right)\right) \right. \right.\nonumber \\
   &\left.\left.-Q^2 r f' \left(4 f'+r^2 h'\right)+8 Q f'
   \left(12 f' h'+r^2 \left(3 \left(h'\right)^2+h\right)+r^2\right)-P^2 r \left(4 f'+r^2 h'\right) \left(f' \left(4 h''+1\right)-4 f'' h'\right)\right) \right.\nonumber \\
   &\left.-4 e^{2 B} f'
   \left(3 f' \left(8 \frac{\partial A}{\partial r} \left(Q \left(4 f'+r^2 h'\right)+h r+r\right)+Q r \left(8 h'+Q r\right)\right)-P^2 r^2 \left(f' \left(4 h''+1\right)-4
   f'' h'\right)\right) \right.\nonumber \\
   &\left.+384 Q \left(f'\right)^3 \frac{\partial A}{\partial r}-e^{6 B} (h+1) r^2 \left(12 f' h'+r^2 \left(3 \left(h'\right)^2+h\right)+r^2\right) \right)
\end{align}

\section{Equations and the Principal Symbol of theory~\eqref{eq:disf_action}}
\label{app:disf_eqns_PS}
The metric equation of motion is given by~\cite{Papallo:2017ddx, Tanahashi:2017kgn}
    \begin{align}
    0&=-\frac{1}{2}\frac{a-bX}{(a-2bX)^2}\mathcal{S}\nabla_{\alpha}\phi\nabla^{\beta}\phi - \frac{1}{2}\frac{X}{a\mathcal{S}}\delta_{\alpha}^{\beta} +\frac{1}{2}\left(\frac{-b}{\mathcal{S}} - 2X\frac{b^2}{a \mathcal{S}^3}\right)\delta_{\alpha \alpha_1 \alpha_2}^{\beta \beta_1 \beta_2} \nabla_{\beta_1}\nabla^{\alpha_1}\phi \nabla_{\beta_2}\nabla^{\alpha_2}\phi \nonumber \\
    &-\frac{1}{2}\frac{b^2}{a \mathcal{S}^3} \delta_{\alpha \alpha_1 \alpha_2 \alpha_3}^{\beta \beta_1 \beta_2 \beta_3}  \nabla_{\beta_1}\nabla^{\alpha_1}\phi \nabla_{\beta_2}\nabla^{\alpha_2}\phi \nabla^{\alpha_3}\phi \nabla_{\beta_3}\phi \nonumber -\frac{1}{4}(a\mathcal{S} + 2X\frac{b}{\mathcal{S}})\delta_{\alpha \alpha_1 \alpha_2}^{\beta \beta_1 \beta_2}R^{\alpha_1 \alpha_2}_{\beta_1 \beta_2} \nonumber \\ 
    &+\frac{1}{4}\delta_{\alpha \alpha_1 \alpha_2 \alpha_3}^{\beta \beta_1 \beta_2 \beta_3}R^{\alpha_1 \alpha_2}_{\beta_1 \beta_2}\nabla^{\alpha_3}\phi\nabla_{\beta_3}\phi,
\end{align}
while the scalar equation is 
    \begin{align}
    0&=-\frac{a-bX}{(a-2bX)^2}\mathcal{S}\Box \phi + \frac{b(2a-bX)}{(a-2bX)^3}\mathcal{S}\nabla_{\alpha_1}\phi\nabla^{\alpha_1}\nabla_{\alpha_2}\phi\nabla^{\alpha_2}\phi +\frac{1}{6}\left(3\frac{b^2}{a\mathcal{S}^3}+12X\frac{b^3}{a^2\mathcal{S}^5}\right)\delta_{\alpha_1 \alpha_2 \alpha_3}^{\beta_1 \beta_2 \beta_3} \nabla^{\alpha_1}\nabla_{\beta_1}\phi \nabla^{\alpha_2}\nabla_{\beta_2}\phi \nabla^{\alpha_3}\nabla_{\beta_3}\phi \nonumber \\
    &+\frac{b^3}{a^2\mathcal{S}^5}\delta_{\alpha_1 \alpha_2 \alpha_3 \alpha_4}^{\beta_1 \beta_2 \beta_3 \beta_4} \nabla^{\alpha_1}\nabla_{\beta_1}\phi \nabla^{\alpha_2}\nabla_{\beta_2}\phi \nabla^{\alpha_3}\nabla_{\beta_3}\phi \nabla^{\alpha_4}\phi \nabla_{\beta_4}\phi +\frac{1}{2}\left(\frac{b}{\mathcal{S}}+2X\frac{b^2}{a\mathcal{S}^3}\right)\delta_{\alpha_1 \alpha_2 \alpha_3}^{\beta_1 \beta_2 \beta_3}\nabla^{\alpha_1}\nabla_{\beta_1}\phi R^{\alpha_2 \alpha_3}_{\beta_2 \beta_3} \nonumber \\
    &+\frac{1}{2}\frac{b^2}{a\mathcal{S}^3}\delta_{\alpha_1 \alpha_2 \alpha_3 \alpha_4}^{\beta_1 \beta_2 \beta_3 \beta_4}\nabla^{\alpha_1}\nabla_{\beta_1}\phi R^{\alpha_2 \alpha_3}_{\beta_2 \beta_3}\nabla^{\alpha_4}\phi\nabla_{\beta_4}\phi,
\end{align}
from which we find the various components of the principal symbol, given by ~\cite{Papallo:2017ddx, Tanahashi:2017kgn} 
    \begin{align}
    (P_{gg}(\xi)\cdot \omega)^{\alpha}{}_{\beta}&= 
    \frac{1}{2}\left(a\mathcal{S}+2X \frac{b}{\mathcal{S}}\right)\delta^{\alpha \alpha_{1}\alpha_{2}}_{\beta \beta_{1} \beta_{2}}\xi_{\alpha_{1}}\xi^{\beta_{1}}\omega_{\alpha_{2}}{}^{\beta_{2}}+\frac{1}{2}\frac{b}{\mathcal{S}}\delta^{\alpha \alpha_{1} \alpha_{2} \alpha_{3}}_{\beta \beta_{1} \beta_{2} \beta_{3}}\xi_{\alpha_{1}}\xi^{\beta_{1}}\omega_{\alpha_{2}}{}^{\beta_{2}}\nabla_{\alpha_{3}}\phi\nabla^{\beta_{3}}\phi, \\[2mm]
    P_{gm}(\xi)^{\alpha}{}_{\beta}&=\left(\frac{-b}{\mathcal{S}}-2X \frac{b^2}{a\mathcal{S}^3}\right)\delta^{\alpha \alpha_{1}\alpha_{2}}_{\beta \beta_{1} \beta_{2}}\xi_{\alpha_{1}}\xi^{\beta_{1}}\nabla_{\alpha_{2}}\nabla^{\beta_{2}}\phi-\frac{b^2}{a\mathcal{S}^3}\delta^{\alpha \alpha_{1}\alpha_{2}\alpha_{3}}_{\beta \beta_{1}\beta_{2}\beta_{3}}\xi_{\alpha_{1}}\xi^{\beta_{1}}\nabla_{\alpha_{2}}\nabla^{\beta_{2}}\phi\nabla_{\alpha_{3}}\phi\nabla^{\beta_{3}}\phi, \\[2mm]
    P_{mm}(\xi)&= -\left(\frac{a-bX}{(a-2bX)^2}\mathcal{S}+2X\frac{b(2a-bX)}{(a-2bX)^3}\mathcal{S}\right)\xi^{2}-\frac{b(2a-bX)}{(a-2bX)^3}\mathcal{S}\delta^{\alpha_{1}\alpha_{2}}_{\beta_{1}\beta_{2}}\xi_{\alpha_{1}}\xi^{\beta_{1}}\nabla_{\alpha_{2}}\phi\nabla^{\beta_{2}}\phi \nonumber\\
    &+\frac{1}{2}\left(\frac{b}{\mathcal{S}}+2X \frac{b^2}{a\mathcal{S}^3}\right)\delta^{\alpha_{1}\alpha_{2}\alpha_{3}}_{\beta_{1}\beta_{2}\beta_{3}} \xi_{\alpha_{1}}\xi^{\beta_{1}}R_{\alpha_{2}\alpha_{3}}{}^{\beta_{2}\beta_{3}}
    +\frac{1}{2}\frac{b^2}{a\mathcal{S}^3}\delta^{\alpha_{1}\alpha_{2}\alpha_{3}\alpha_{4}}_{\beta_{1}\beta_{2}\beta_{3}\beta_{4}}\xi_{\alpha_{1}}\xi^{\beta_{1}}\nabla_{\alpha_{2}}\phi\nabla^{\beta_{2}}\phi R_{\alpha_{3}\alpha_{4}}{}^{\beta_{3}\beta_{4}}\nonumber\\
    &+\left(3\frac{b^2}{a\mathcal{S}^3}+2X\frac{3b^3}{a^2\mathcal{S}^5}\right)\delta^{\alpha_{1}\alpha_{2}\alpha_{3}}_{\beta_{1}\beta_{2}\beta_{3}}\xi_{\alpha_{1}}\xi^{\beta_{1}}\nabla_{\alpha_{2}}\nabla^{\beta_{2}}\phi\nabla_{\alpha_{3}}\nabla^{\beta_{3}}\phi\nonumber\\
    &+\frac{3b^3}{a^2\mathcal{S}^5}\delta^{\alpha_{1}\alpha_{2}\alpha_{3}\alpha_{4}}_{\beta_{1}\beta_{2}\beta_{3}\beta_{4}}\xi_{\alpha_{1}}\xi^{\beta_{1}}\nabla_{\alpha_{2}}\nabla^{\beta_{2}}\phi\nabla_{\alpha_{3}}\nabla^{\beta_{3}}\phi\nabla_{\alpha_{4}}\phi\nabla^{\beta_{4}}\phi.
\end{align}
\end{widetext}
\clearpage
\bibliography{biblio.bib}
\end{document}